# Indium Hydroxide Ceramic Targets: A Breakthrough in High-Mobility Thin-Film Transistor Technology

Hikaru Sadahira[1,#] Prashant R. Ghediya[2,#] Hyeonjun Kong[1,#] Akira Miura[3], Yasutaka Matsuo[2], Hiromichi Ohta[2,*] and Yusaku Magari[2,*]

[1] *Graduate School of Information Science and Technology, Hokkaido University, N14W9, Sapporo 060-0814, Japan*

[2] *Research Institute for Electronic Science, Hokkaido University, N20W10, Sapporo 001-0020, Japan*

[3] *Graduate School of Engineering, Hokkaido University, N13W8, Kita, Sapporo 060-8628, Japan*

*Corresponding author. hiromichi.ohta@es.hokudai.ac.jp, magari.yusaku@es.hokudai.ac.jp
#Equally contributed to this work

**ABSTRACT**: Thin-film transistors composed of a hydrogen-containing indium oxide active layer are promising candidates for backplane devices in next-generation flat panel displays, offering higher definition and faster operation. However, the hydrogen incorporation process during film deposition poses challenges for scalable and industrial development due to both safety and controllability issues. Here, we demonstrate that using indium hydroxide ceramic as the target material for film deposition overcomes the difficulties associated with hydrogen gas usage. We sintered commercially available indium hydroxide powder using a conventional ceramic process at 150 – 250 °C in air and utilized it for the deposition of hydrogen-incorporated indium oxide films via pulsed laser deposition. The resulting indium oxide films, after thermal annealing, contained a sufficient concentration of hydrogen and exhibited very high electron mobility due to significantly grown grains. Furthermore, we confirmed that the fabricated thin-film transistors exhibited comparably high performance to those produced using the gas-phase hydrogen incorporation method. This approach offers a practical pathway for hydrogen-containing indium oxide-based thin-film transistors in next-generation flat panel displays.

**INTRODUCTION**

Thin-film transistors (TFTs) are crucial for controlling pixel brightness in flat-panel displays (FPDs), such as liquid crystal displays (LCDs) and organic light-emitting diodes (OLEDs).[1-3] Currently, amorphous InGaZnO$_4$ (IGZO)-based TFTs are widely commercialized for FPDs.[4-6] However, their field-effect mobility ($\mu_{FE}$) is typically limited to $5 - 10$ cm$^2$ V$^{-1}$ s$^{-1}$,[2, 7, 8] which is insufficient for next-generation displays requiring higher resolution (8 K) and faster refresh rates (240 Hz).[5, 9, 10] To address these limitations, extensive efforts have been made to develop oxide semiconductor-based TFTs with enhanced $\mu_{FE}$.[11-23]

Among various oxide semiconductors, polycrystalline indium oxide (In$_2$O$_3$) has emerged as a promising candidate due to its large lateral grain size and superior carrier transport properties. Previous studies have shown that incorporating hydrogen during In$_2$O$_3$ film deposition suppresses premature crystallization, enabling large-grain growth upon thermal annealing at ~250 °C.[24] This approach has led to high-performance TFTs with $\mu_{FE}$ values reaching ~140 cm$^2$ V$^{-1}$ s$^{-1}$.[25] However, these devices suffer from bias stress instability, which has been mitigated through surface passivation with yttrium oxide (Y$_2$O$_3$) and erbium oxide (Er$_2$O$_3$).[26, 27]

Despite these advancements, a major challenge remains in the mass production of high-mobility In$_2$O$_3$-based TFTs. Current hydrogen incorporation methods rely on complex and hazardous techniques, such as the introducing hydrogen gas[25, 28-30] or manipulating deposition chamber pressure to utilize water vapor.[24] These methods pose safety risks and operational challenges, making them unsuitable for large-scale manufacturing. Therefore, a simpler, safer, and more efficient hydrogen introduction method is essential for the commercial viability of In$_2$O$_3$-based TFTs.

In this study, we propose a novel and safe approach using indium hydroxide (In(OH)$_3$) ceramic targets for film deposition (**Figure 1**). In(OH)$_3$ inherently contains hydrogen and undergoes thermal decomposition at ~200 °C, transforming into In$_2$O$_3$. This property makes it an ideal hydrogen source during deposition. By utilizing In(OH)$_3$ as a target material, hydrogen is directly incorporated into the as-deposited In$_2$O$_3$ films without the need of external hydrogen sources, significantly simplifying the process. Our results demonstarate that films deposited using In(OH)$_3$ ceramic targets exhibit a high hydrogen centent (~$10^{22}$ cm$^{-3}$), promoting large lateral grain growth upon annealing. The resultant TFTs display excellent switching behavior and a high $\mu_{FE}$ of ~81 cm$^2$ V$^{-1}$ s$^{-1}$. This breakthrough brings In$_2$O$_3$-based TFTs closer to commercial viability by providing a scalable and safer route for next-generation display applications.

**EXPERIMENTAL SECTION**

**Fabrication of In(OH)$_3$ ceramic targets**: High-purity In(OH)$_3$ powder (purity >92%, FUJIFILM Wako Pure Chemical Co.) was uniaxially pressed into pellets (18 mm in diameter, ~5 mm in thickness). The pellets were then subjected to cold isostatic pressing (CIP) at 40 MPa for 1 min, followed by sintering in air at 150 – 250 °C for 24 h using a muffle furnace. The DSC and TG-DTA of In(OH)$_3$ were conducted to determine the optimal sintering temperature. The relative density and Vickers hardness of the In(OH)$_3$ targets were measured. Crystallographic phases of the sintered targets were analyzed by a powder X-ray diffraction (XRD, Miniflex600, Rigaku Co).

**Fabrication of thin films using the In(OH)$_3$ ceramic targets**: The resultant In(OH)$_3$ ceramic target was placed in the PLD chamber, which was evacuated to a base pressure below 1 × 10$^{-4}$ Pa. Focused KrF excimer laser pulses (~1.5 J cm$^{-2}$ pulse$^{-1}$, 10 Hz) were irradiated onto the In(OH)$_3$ ceramic target. The oxygen pressure during deposition was maintained at 3 Pa, and 50-nm-thick films were deposited on alkali-free glass substrates (EAGLE XG®, Corning®). After deposition, the films were thermally annealed at 300 °C for 0.5 h in air.

**Characterization of the resultant films**: Film thickness, roughness, and crystallographic phases were analyzed using high-resolution XRD (ATX-G, Rigaku Co.) with monochromated Cu K$\alpha_1$ ($\lambda$ = 0.154059 nm) radiation. The incident angle ($\omega$) was fixed at 0.5°, and 2$\theta$ scanning was performed to obtain XRD patterns. Microstructure and roughness were observed using atomic force microscopy (AFM, Nanocute, Hitachi Hi-Tech Sci. Co.). Crystallographic orientations and lateral grain size were investigated using electron backscatter diffraction (EBSD, EDAX-TSL Hikari High Speed EBSD Detector). The carrier concentrations ($n$) and Hall mobility ($\mu_{Hall}$) were measured using laboratory-made Hall effect measurements in the van der Pauw configuration under a ±7600 G magnetic field at room temperature. Thermopower ($S$) was measured using a steady-state method at room temperature. The hydrogen and carbon concentrations in the films was analyzed using secondary ion mass spectrometry (SIMS, ULVAC-PHI, ADEPT-1010) with $Cs^+$ as a primary ion.

**Fabrication of thin-film transistors**: Bottom-gate, top-contact TFTs were fabricated on ITO-coated alkali-free glass substrates. A 100-nm-thick amorphous $Al_2O_3$ film, with a dielectric permittivity ($\varepsilon_r$) of 8 and capacitance ($C_i$) of 70.8 nF cm$^{-2}$, was deposited by an atomic layer deposition technique (ALD, R-200 Advanced, Picosun Oy) as the gate insulator. 5-nm-thick films $In_2O_3$ films were deposited by PLD using a stencil mask without substrate heating from a ceramic $In(OH)_3$ target sintered at 150 − 250 °C. The oxygen pressure during film deposition was maintained at 3 Pa. The multilayer film was then annealed at 300 °C for 0.5 h in air. 100-nm-thick ITO films were deposited by PLD through a stencil mask as the source and drain electrodes. Subsequently, a $Y_2O_3$ film was deposited by PLD as a passivation layer. Finally, the TFTs were annealed at 350 °C for 0.5 h in air. The design channel length ($L$) and the channel width ($W$) were 200 and 400 μm, respectively.

**Transistor characteristics measurements**: The transfer characteristics of the resultant $In_2O_3$ TFTs were measured using a semiconductor device analyzer (B1500A, Agilent) in the dark. The field effect mobility ($\mu_{FE}$) was calculated from the linear transfer characteristics at a drain voltage ($V_d$) of 5 V using equation (1).

$$\mu_{FE} = \frac{g_m}{C_i \frac{W}{L} V_d}, \qquad (1)$$

where $g_m$ is the transconductance, $C_i$ is the oxide capacitance of the gate insulator (70.8 nF cm$^{-2}$ for amorphous $Al_2O_3$), and $V_d$ is the drain voltage. The subthreshold swing was extracted from $V_g$, which required an increase in the drain current ($I_d$) from 1 to 10 pA.

Positive gate bias stress (PBS, $V_g$ = +20 V) and negative gate bias stress (NBS, $V_g$ = −20 V) tests were conducted for $In_2O_3$ TFTs for 5000 s at room temperature in air ambient.

**RESULTS AND DISCUSSION**

We first fabricated $In(OH)_3$ ceramic targets. Since no prior reports exist on $In(OH)_3$ ceramic targets, it was necessary to determine an appropiate sintering temperature. To address this, we conducted differential scanning calorimetry (DSC) and thermogravimetric-differential thermal analysis (TG-DTA) on $In(OH)_3$ powder (**Figure 2a**). The DSC and TG-DTA curves revealed that a small endothermic DTA peak around 90 °C, accompanied by slight weight loss, attributed to the vaporization of surface-adsorbed water molecules. A broader endothermic peak around 230 °C, with gradual weight loss, corresponds to the dehydration of $In(OH)_3$. An abrupt exothermic peak at approximately 250 °C indicates the crystallization of $In_2O_3$. Based on these results, we selected a sintering temperature range of 150 to 250 °C for $In(OH)_3$ ceramic targets. To examine the effects of sintering temperature, we prepared compacted $In(OH)_3$ ceramic targets (18 mm in diameter, 5 mm in thickness) and sintered them at 150, 180, 200, and 250 °C for 24 hours in air.

**Figure 2b** shows the powder X-ray diffraction (XRD) patterns of the resultant $In(OH)_3$ ceramic targets. Below 180 °C, the $In(OH)_3$ and $InOOH$ phases remained stable. Above 200 °C, diffraction peaks corresponding to $In_2O_3$ began to appear, and their intensity increases with temperature, indicating crystal growth of $In_2O_3$. These results suggest that by varying the sintering temperature of $In(OH)_3$, the hydrogen concentration in the ceramic target can be adjusted, enabling controlled hydrogen incorporation into the $In_2O_3$ films.

We next evaluated the mechanical properties of the $In(OH)_3$ ceramic targets. The target color changed from white to light yellow above 200 °C (**Figure 3a**), consistent with the phase transition from $In(OH)_3$ (white) to $In_2O_3$ (light yellow). However, no noticeable changes in overall dimensions were observed. High-temperature scanning electron microscopy (SEM) (**Figure S1**) further confirmed that no significant morphological changes occurred up to 250 °C, supporting the observation that the overall size remained unchanged within this temperature range. However, when sintering temperatures exceeded 200 °C, the relative density of the targets significantly decreased. Similarly, Vickers hardness measurements indicated a notable reduction under these conditions. Optical micrographs of the ceramic targets (**Figure 3b**) revealed an increase in the number of open pores with higher sintering temperatures, while SEM observations (**Figure 3c**) suggested that grain size remained largely unchanged.

These results indicate that dehydration of $In(OH)_3$ and crystallization of $In_2O_3$ occur above 200 °C. However, densification of $In_2O_3$ does not proceed at these temperatures. Consequently, $In(OH)_3$ ceramic targets sintered above 200 °C exhibit relatively low density and poor mechanical properties.

Using the In(OH)$_3$ ceramic targets, we deposited ~50-nm-thick films on alkali-free glass substrates by pulsed laser deposition (PLD) at room temperature. Details of the deposition process are provided in the **Experimental Section**.

To investigate hydrogen incorporation into the films, secondary ion mass spectrometry (SIMS) measurements were performed (**Figure 4a**). The as-deposited films from the 150 °C-sintered In(OH)$_3$ ceramic target contained a high hydrogen concentration of $1 \times 10^{22}$ cm$^{-3}$. Given that the hydrogen concentration in the 150 °C-sintered, low-density In(OH)$_3$ target (2.01 g/cm$^3$) is calculated to be $2 \times 10^{22}$ cm$^{-3}$, this result indicates that nearly all the hydrogen from the In(OH)$_3$ ceramic target was successfully incorporated into the In$_2$O$_3$ film. Increasing the sintering temperature of In(OH)$_3$ to 200 °C resulted in a slight decreas in hydrogen concentration to $5 \times 10^{21}$ cm$^{-3}$. Although thermal decomposition of In(OH)$_3$ into In$_2$O$_3$ occurred after sintering at 200 °C (**Figure 2b**), a relatively large amount of hydrogen remained in the In$_2$O$_3$ film. This is likely due to moisture retention within the target, as the 200 °C-sintered In(OH)$_3$ ceramic target exhibited a porous structure and low density (**Figure 3**).

In contrast, films deposited from a conventional In$_2$O$_3$ ceramic target contained only ~$3 \times 10^{20}$ cm$^{-3}$ hydrogen, which originated from residual hydrogen in the deposition chamber.[31] These results confirm that In(OH)$_3$ ceramic targets effectively introduce hydrogen into the films during deposition. We also measured carbon impurities using SIMS and found that the carbon concentration in the In$_2$O$_3$ films was lower when using the In(OH)$_3$ ceramic target compared to the conventional dense In$_2$O$_3$ ceramic target (**Figure S2**).

We then investigated the electron transport properties of the as-deposited films at room temperature (**Figure 4c**). The carrier concentration ranged from $5 \times 10^{20}$ to $7 \times 10^{20}$ cm$^{-3}$, consistent with the observed thermopower values ($-S = 7 - 10$ μV K$^{-1}$). The Hall mobility ($\mu_{\text{Hall}}$) ranged from 10 to 20 cm$^2$ V$^{-1}$ s$^{-1}$. These results suggest that hydrogen in the as-deposited films exists as protons, significantly influencing electron transport properties. The films were thermally annealed at 300 °C for 0.5 h in air. SIMS results (**Figure 4b**) revealed that the hydrogen concentration in films deposited from an In$_2$O$_3$ ceramic target remained unchanged after annealing at 300 °C. In contrast, the hydrogen concentration in films deposited from the In(OH)$_3$ ceramic targets slightly decreased after annealing but remained approximately one order of magnitude higher than that in films from the In$_2$O$_3$ ceramic target.

A comparison between the as-deposited and annealed films is shown in **Figure S3**. Despite retaining a large amount of hydrogen ($3 - 5 \times 10^{21}$ cm$^{-3}$), the carrier concentration (**Figure 4c**) significantly decreased to ~$10^{17}$ cm$^{-3}$ after annealing. The absolute value of thermopower increased correspondingly ($-S = 240 - 300$ μV K$^{-1}$), reflecting the reduction in carrier concentration. These results suggest that the decrease in carrier concentration after annealing is primarily due to a reduction in oxygen vacancies rather than hydrogen desorption. A similar phenomenon has been reported for $In_2O_3$ films deposited with hydrogen gas.[25] Importantly, the Hall mobility of the annealed films remained in the range of $10 - 20$ cm$^2$ V$^{-1}$ s$^{-1}$, indicating that mobility was not significantly affected by the annealing process.

To analyze the structure of the deposited films, grazing incidence XRD patterns were measured (**Figure 5**) at an incident X-ray angle of 0.5°. The as-deposited films from $In(OH)_3$ ceramic targets exhibited only halo patterns around $2\theta = 32°$, indicating an amorphous structure (**Figure 5a**). In contrast, films deposited from dense $In_2O_3$ ceramic targets displayed intense diffraction peaks, suggesting a randomly oriented polycrystalline structure. After annealing at 300 °C, all films exhibited intense diffraction peaks, confirming the formation of polycrystalline structures (**Figure 5b**).

We further analyzed the lateral grain size of the films using electron backscatter diffraction (EBSD) images (**Figure 6**). As-deposited films from $In(OH)_3$ ceramic targets remained amorphous (**Figure 6a**), while large grains appeared after annealing at 300 °C (**Figure 6b**). Notably, the largest grain size of approximately 2 μm was observed in films deposited using an $In(OH)_3$ ceramic target sintered at 200 °C (**Figure 6c**). In contrast, films deposited from dense $In_2O_3$ ceramic target exhibited a smaller grain structure (0.23 μm) in the as-deposited state, with no significant grain growth after annealing at 300 °C.

These results indicate that using $In(OH)_3$ ceramic targets supprese the crystallization of $In_2O_3$ during deposition due to hydrogen incorporation, while promoting singnificant grain growth upon annealing. However, sintering at temperatures above the thermal decomposition temperature of $In(OH)_3$ promotes its decomposition into $In_2O_3$ (**Figs. 2 and 3**), which limits the incorporation of sufficient hydrogen into the $In_2O_3$ films. As a result, when using the $In(OH)_3$ target sintered at 250 °C, dehydration and crystallization occurred at lower temperatures, leading to smaller crystal grains. X-ray reflectivity (XRR) measurements confirmed a slight increase in film density during the crystallization process (**Figure S4**). Regarding optical properties, a decrease in subgap tail state associated with the amorphous structure was observed after crystallization, leading to a slight increase in the optical bandgap (**Figure S5**). Additionally, atomic force microscopy (AFM) observations revealed that the root mean square roughness ($R_{rms}$) of the large-grain $In_2O_3$ film was approximately 0.25 nm (**Figure S6**), indicating an extremely smooth surface, which is crucial for device applications.

Finally, we fabricated TFTs using the $In(OH)_3$ ceramic target to deposit active $In_2O_3$ channel layer (**Figure 7a**). Details of the TFT fabrication are described in the **Experimental Section**. The channel length (*L*) and channel width (*W*), corrected for patterning errors caused by film deposition through the stencil mask, were used in the field effect mobility calculation.

**Figure S7** shows output characteristics of the $In_2O_3$ TFTs with the channel deposited using an $In(OH)_3$ ceramic target sintered at 200 °C. The drain current ($I_d$) showed pinch-off behavior and current saturation characteristics at higher drain voltage ($V_d$). The transfer characteristics of the resultant TFTs were measured at room temperature with a $V_d$ was 5 V, while the gate voltage ($V_g$) was swept from −10 V to +20 V. The TFT fabricated using an $In_2O_3$ ceramic target did not show clear switching-off behavior (**Figure 7b**) due to the excessively high residual carrier concentration in the $In_2O_3$ channel ($n \sim 10^{19}$ cm$^{-3}$). In contrast, TFTs fabricated using the $In(OH)_3$ ceramic target exhibited excellent switching behavior with a high $\mu_{FE}$ of >60 cm$^2$ V$^{-1}$ s$^{-1}$ (**Figure 7c**). Notably, the TFT fabricated using the 200 °C-annealed $In(OH)_3$ ceramic target demonstrated the highest $\mu_{FE}$ of ~81 cm$^2$ V$^{-1}$ s$^{-1}$, likely due to the largest grain size among the samples. We further evaluated the reliability of the $In_2O_3$ TFTs. As shown in **Figure S8**, the TFTs exhibited excellent stability under both positive gate-bias stress (PBS; $V_g$ = +20 V) and negative gate-bias stress (NBS; $V_g$ = −20 V). The threshold voltage shift of the $In_2O_3$ TFT after 5000 s was 0.51 V under PBS and 0.17 V under NBS at room temperature in air.

Compared to previous methods for introducing hydrogen into as-deposited $In_2O_3$ films, present approach offers a significant advantage. Since the $In(OH)_3$ ceramic target inherently contains sufficient hydrogen, external hydrogen gas introduction is unnecessary. This eliminates the need for gas cylinders and precise control of the pressure in the deposition chamber, making the process both safer and more straightforward. These advantages make the proposed method highly promising for practical applications in next-generation display technologies.

While the current method offers several advantages, there is still room for improvement. Specifically, the relatively low bulk density of the In(OH)$_3$ ceramic target, resulting from pressure-less sintering at lower temperatures, has led to lower Vickers hardness. To address this, we plan to explore advanced densification techniques, such as hot pressing ,to increase the bulk density of the ceramic target. This improvement will enhance both the mechanical properties of the ceramic target and the overall performance of the thin films.

**CONCLUSIONS**

In summary, we proposed a safe and simple method for hydrogen incorporation into In$_2$O$_3$ films using an In(OH)$_3$ ceramic target. The In(OH)$_3$ ceramic target was fabricated using conventional ceramic processing, and thin films were deposited at room temperature by pulsed laser deposition. The as-deposited films exhibited a hydrogen concentration of ~$10^{22}$ cm$^{-3}$ when the In(OH)$_3$ ceramic target was used, compared to only ~$10^{20}$ cm$^{-3}$ when a conventional In$_2$O$_3$ ceramic target was employed. Upon annealing, significant grain growth was observed in films containing ~$10^{22}$ cm$^{-3}$ hydrogen, whereas no such grain growth occurred in films with ~$10^{20}$ cm$^{-3}$ hydrogen. The resultant TFTs, fabricated using the In(OH)$_3$ ceramic target, exhibited a high field-effect mobility of ~81 cm$^2$ V$^{-1}$ s$^{-1}$. The present approach advances In$_2$O$_3$-based TFTs technology toward practical applications in next-generation flat-panel displays, enabling large-area fabrication and high refresh rate capabilities.

**AUTHOR INFORMATION**

**Corresponding Authors**

**Hiromichi Ohta** − Research Institute for Electronic Science, Hokkaido University, N20W10, Kita, Sapporo 001-0020, Japan

ORCID: orcid.org/0000-0001-7013-0343

Email: hiromichi.ohta@es.hokudai.ac.jp

**Yusaku Magari** − Research Institute for Electronic Science, Hokkaido University, N20W10, Kita, Sapporo 001-0020, Japan

ORCID: orcid.org/0000-0001-9655-4283

Email: magari.yusaku@es.hokudai.ac.jp

**Author**

**Hikaru Sadahira** − Graduate School of Information Science and Technology, Hokkaido University. N14W9, Kita Sapporo 060-0814, Japan
ORCID: orcid.org/0009-0005-5925-0922

**Prashant R. Ghediya** − Research Institute for Electronic Science, Hokkaido University, N20W10, Kita, Sapporo 001-0020, Japan
ORCID: orcid.org/0000-0001-9953-0471

**Hyeonjun Kong** − Graduate School of Information Science and Technology, Hokkaido University. N14W9, Kita Sapporo 060-0814, Japan
ORCID: orcid.org/0009-0004-0358-2162

**Akira Miura** − Graduate School of Engineering, Hokkaido University, N13W8, Kita, Sapporo 060-8628, Japan
ORCID: orcid.org/0000-0003-0388-9696

**Yasutaka Matsuo** − Research Institute for Electronic Science, Hokkaido University, N20W10, Kita, Sapporo 001-0020, Japan
ORCiD: orcid.org/0000-0002-5071-0284

**Author Contributions**

H.S., P.R.G. and H.K. fabricated the ceramic targets, thin films, and transistors, and performed their characterization. M.A. performed TG-DTA measurements and high-temperature SEM measurements. Y. Matsuo fabricated $AlO_x$ gate insulator films. H.O. provided the idea of use of $In(OH)_3$ target. H.O. and Y. Magari planned and supervised the project. All the authors discussed the results and commented on the manuscript.

**ACKNOWLEDGEMENTS**

We thank Y. Miyakoshi for their support in the Vickers hardness measurements. This work was supported by Japan Society for the Promotion of Science (JSPS) KAKENHI Grant Number 22K14303 (to Y. Magari). A.M. was supported by JST-PRESTO. H.O. was supported by Grant-in-Aid for Scientific Research A from JSPS (22H00253). H.K. was supported by a Hokkaido University EXEX Doctoral Fellowship (JPMJSP2119). Part of this work was supported by the Advanced Research Infrastructure for Materials and Nanotechnology in Japan (JPMXP1224HK0082, JPMXP1224AT0216) of the Ministry of Education, Culture, Sports, Science, and Technology (MEXT). Part of this work was supported by the Crossover Alliance to Create the Future with People, Intelligence and Materials, and by the Network Joint Research Center for Materials and Devices.

**Notes**

The authors declare no competing interest.

**REFERENCES**

(1) Hirao, T.; Furuta, M.; Furuta, H.; Matsuda, T.; Hiramatsu, T.; Hokari, H.; Yoshida, M.; Ishii, H.; Kakegawa, M. Novel top-gate zinc oxide thin-film transistors (ZnO TFTs) for AMLCDs. *Journal of the Society for Information Display* **2007**, *15* (1), 17-22. DOI: 10.1889/1.2451545.

(2) Kamiya, T.; Hosono, H. Material characteristics and applications of transparent amorphous oxide semiconductors. *NPG Asia Materials* **2010**, *2* (1), 15-22. DOI: 10.1038/asiamat.2010.5.

(3) Park, J. S.; Lim, J. H. High Mobility Oxide Thin-film Transistors for AMOLED Displays. *SID Symposium Digest of Technical Papers* **2022**, *53* (1), 20-23. DOI: 10.1002/sdtp.15405.

(4) Hosono, H.; Kumomi, H. Amorphous Oxide Semiconductors: IGZO and Related Materials for Display and Memory. **2022**, Book.

(5) Yeon Kwon, J.; Kyeong Jeong, J. Recent progress in high performance and reliable n-type transition metal oxide-based thin film transistors. *Semiconductor Science and Technology* **2015**, *30* (2). DOI: 10.1088/0268-1242/30/2/024002.

(6) Takeda, Y.; Aman, M.; Murashige, S.; Ito, K.; Ishida, I.; Matsukizono, H.; Makita, N. Automotive OLED Display with High Mobility Top Gate IGZO TFT Backplane. *ITE Transactions on Media Technology and Applications* **2020**, *8* (4), 224-229. DOI: 10.3169/mta.8.224.

(7) Nomura, K.; Ohta, H.; Takagi, A.; Kamiya, T.; Hirano, M.; Hosono, H. Room-temperature fabrication of transparent flexible thin-film transistors using amorphous oxide semiconductors. *Nature* **2004**, *432* (7016), 488-492. DOI: 10.1038/nature03090.

(8) Nomura, K.; Takagi, A.; Kamiya, T.; Ohta, H.; Hirano, M.; Hosono, H. Amorphous Oxide Semiconductors for High-Performance Flexible Thin-Film Transistors. *Japanese Journal of Applied Physics* **2006**, *45* (5S). DOI: 10.1143/jjap.45.4303.

(9) Kamiya, T.; Nomura, K.; Hosono, H. Present status of amorphous In-Ga-Zn-O thin-film

transistors. *Sci Technol Adv Mater* **2010**, *11* (4), 044305. DOI: 10.1088/1468-6996/11/4/044305.

(10) Hara, Y.; Kikuchi, T.; Kitagawa, H.; Morinaga, J.; Ohgami, H.; Imai, H.; Daitoh, T.; Matsuo, T. IGZO-TFT technology for large-screen 8K display. *Journal of the Society for Information Display* **2018**, *26* (3), 169-177. DOI: 10.1002/jsid.648.

(11) Ebata, K.; Tomai, S.; Tsuruma, Y.; Iitsuka, T.; Matsuzaki, S.; Yano, K. High-Mobility Thin-Film Transistors with Polycrystalline In–Ga–O Channel Fabricated by DC Magnetron Sputtering. *Applied Physics Express* **2012**, *5* (1). DOI: 10.1143/apex.5.011102.

(12) Cho, S. H.; Ko, J. B.; Ryu, M. K.; Yang, J.-H.; Yeom, H.-I.; Lim, S. K.; Hwang, C.-S.; Park, S.-H. K. Highly Stable, High Mobility Al:SnZnInO Back-Channel Etch Thin-Film Transistor Fabricated Using PAN-Based Wet Etchant for Source and Drain Patterning. *IEEE Transactions on Electron Devices* **2015**, *62* (11), 3653-3657. DOI: 10.1109/ted.2015.2479592.

(13) Sheng, J.; Hong, T.; Lee, H. M.; Kim, K.; Sasase, M.; Kim, J.; Hosono, H.; Park, J. S. Amorphous IGZO TFT with High Mobility of approximately 70 $cm^2$/(Vs) via Vertical Dimension Control Using PEALD. *ACS Appl Mater Interfaces* **2019**, *11* (43), 40300-40309. DOI: 10.1021/acsami.9b14310.

(14) Shiah, Y.-S.; Sim, K.; Shi, Y.; Abe, K.; Ueda, S.; Sasase, M.; Kim, J.; Hosono, H. Mobility–stability trade-off in oxide thin-film transistors. *Nature Electronics* **2021**, *4* (11), 800-807. DOI: 10.1038/s41928-021-00671-0.

(15) Kim, B. K.; On, N.; Choi, C. H.; Kim, M. J.; Kang, S.; Lim, J. H.; Jeong, J. K. Polycrystalline Indium Gallium Tin Oxide Thin-Film Transistors With High Mobility Exceeding 100 $cm^2$/Vs. *IEEE Electron Device Letters* **2021**, *42* (3), 347-350. DOI: 10.1109/led.2021.3055940.

(16) Cho, M. H.; Choi, C. H.; Seul, H. J.; Cho, H. C.; Jeong, J. K. Achieving a Low-Voltage, High-Mobility IGZO Transistor through an ALD-Derived Bilayer Channel and a Hafnia-Based Gate Dielectric Stack. *ACS Appl Mater Interfaces* **2021**, *13* (14), 16628-16640. DOI: 10.1021/acsami.0c22677.

(17) Shiah, Y.-S.; Sim, K.; Ueda, S.; Kim, J.; Hosono, H. Unintended Carbon-Related Impurity and Negative Bias Instability in High-Mobility Oxide TFTs. *IEEE Electron Device Letters* **2021**, *42* (9), 1319-1322. DOI: 10.1109/led.2021.3101654.

(18) Lee, J.; Choi, C. H.; Kim, T.; Hur, J.; Kim, M. J.; Kim, E. H.; Lim, J. H.; Kang, Y.; Jeong, J. K. Hydrogen-Doping-Enabled Boosting of the Carrier Mobility and Stability in Amorphous IGZTO Transistors. *ACS Appl Mater Interfaces* **2022**, *14* (51), 57016-57027. DOI: 10.1021/acsami.2c18094.

(19) Rabbi, M. H.; Lee, S.; Sasaki, D.; Kawashima, E.; Tsuruma, Y.; Jang, J. Polycrystalline

InGaO Thin-Film Transistors with Coplanar Structure Exhibiting Average Mobility of approximately 78 cm$^2$ V$^{-1}$ s$^{-1}$ and Excellent Stability for Replacing Current Poly-Si Thin-Film Transistors for Organic Light-Emitting Diode Displays. *Small Methods* **2022**, *6* (9), e2200668. DOI: 10.1002/smtd.202200668.

(20) Yang, H.; Zhang, Y.; Matsuo, Y.; Magari, Y.; Ohta, H. Thermopower Modulation Analyses of High-Mobility Transparent Amorphous Oxide Semiconductor Thin-Film Transistors. *ACS Applied Electronic Materials* **2022**, *4* (10), 5081-5086. DOI: 10.1021/acsaelm.2c01210.

(21) Kim, G. B.; On, N.; Kim, T.; Choi, C. H.; Hur, J. S.; Lim, J. H.; Jeong, J. K. High Mobility IZTO Thin-Film Transistors Based on Spinel Phase Formation at Low Temperature through a Catalytic Chemical Reaction. *Small Methods* **2023**, *7* (7), e2201522. DOI: 10.1002/smtd.202201522.

(22) Kim, T.; Choi, C. H.; Hur, J. S.; Ha, D.; Kuh, B. J.; Kim, Y.; Cho, M. H.; Kim, S.; Jeong, J. K. Progress, Challenges, and Opportunities in Oxide Semiconductor Devices: A Key Building Block for Applications Ranging from Display Backplanes to 3D Integrated Semiconductor Chips. *Adv Mater* **2023**, *35* (43), e2204663. DOI: 10.1002/adma.202204663.

(23) Wen, P.; Peng, C.; Chen, Z.; Ding, X.; Chen, F.-H.; Yan, G.; Xu, L.; Wang, D.; Sun, X.; Chen, L.; et al. High mobility of IGO/IGZO double-channel thin-film transistors by atomic layer deposition. *Applied Physics Letters* **2024**, *124* (13). DOI: 10.1063/5.0194691.

(24) Koida, T.; Fujiwara, H.; Kondo, M. Hydrogen-doped In$_2$O$_3$ as High-mobility Transparent Conductive Oxide. *Japanese Journal of Applied Physics* **2007**, *46* (7L). DOI: 10.1143/jjap.46.L685.

(25) Magari, Y.; Kataoka, T.; Yeh, W.; Furuta, M. High-mobility hydrogenated polycrystalline In$_2$O$_3$ (In$_2$O$_3$:H) thin-film transistors. *Nat Commun* **2022**, *13* (1), 1078. DOI: 10.1038/s41467-022-28480-9.

(26) Ghediya, P. R.; Magari, Y.; Sadahira, H.; Endo, T.; Furuta, M.; Zhang, Y.; Matsuo, Y.; Ohta, H. Reliable Operation in High-Mobility Indium Oxide Thin Film Transistors. *Small Methods* **2025**, *9* (1), e2400578. DOI: 10.1002/smtd.202400578   From NLM PubMed-not-MEDLINE.

(27) Ohta, H.; Kambayashi, T.; Nomura, K.; Hirano, M.; Ishikawa, K.; Takezoe, H.; Hosono, H. Transparent Organic Thin-Film Transistor with a Laterally Grown Non-Planar Phthalocyanine Channel. *Advanced Materials* **2004**, *16* (4), 312-316. DOI: 10.1002/adma.200306015.

(28) Furuta, M.; Shimpo, K.; Kataoka, T.; Tanaka, D.; Matsumura, T.; Magari, Y.; Velichko, R.; Sasaki, D.; Kawashima, E.; Tsuruma, Y. High Mobility Hydrogenated Polycrystalline In-Ga-O (IGO:H) Thin-Film Transistors formed by Solid Phase Crystallization. *SID Symposium Digest*

*of Technical Papers* **2021**, *52* (1), 69-72. DOI: 10.1002/sdtp.14612.

(29) Wang, X.; Magari, Y.; Furuta, M. Nucleation and grain growth in low-temperature rapid solid-phase crystallization of hydrogen-doped indium oxide. *Japanese Journal of Applied Physics* **2024**, *63* (3). DOI: 10.35848/1347-4065/ad21ba.

(30) Okamoto, N.; Wang, X.; Morita, K.; Kato, Y.; Alom, M. M.; Magari, Y.; Furuta, M. Uniformity and Reliability of Enhancement-Mode Polycrystalline Indium Oxide Thin Film Transistors Formed by Solid-Phase Crystallization. *IEEE Electron Device Letters* **2024**, *45* (12), 2403-2406. DOI: 10.1109/led.2024.3480991.

(31) Miyase, T.; Watanabe, K.; Sakaguchi, I.; Ohashi, N.; Domen, K.; Nomura, K.; Hiramatsu, H.; Kumomi, H.; Hosono, H.; Kamiya, T. Roles of Hydrogen in Amorphous Oxide Semiconductor In-Ga-Zn-O: Comparison of Conventional and Ultra-High-Vacuum Sputtering. *ECS Journal of Solid State Science and Technology* **2014**, *3* (9), Q3085-Q3090. DOI: 10.1149/2.015409jss.

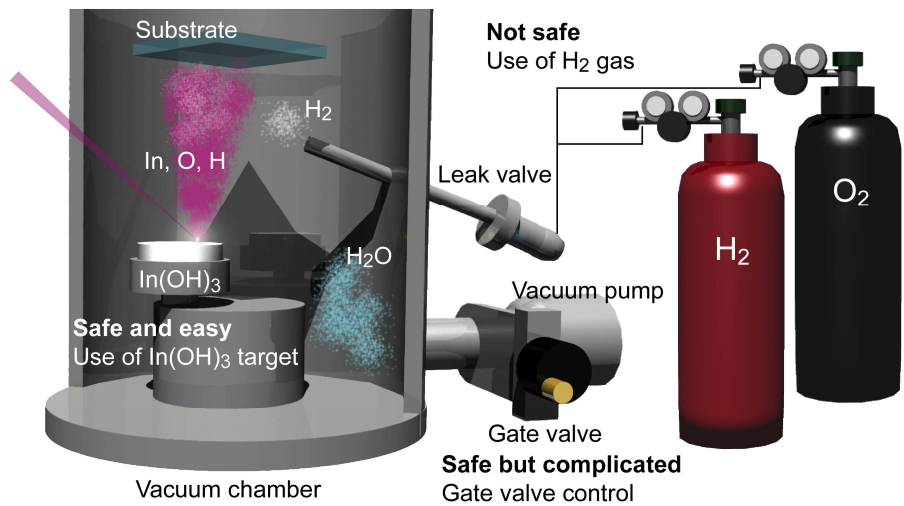

**Figure 1. Hydrogen introduction into polycrystalline In$_2$O$_3$ thin films composed of large grains.** Schematic illustration of the three different hydrogen introduction methods during film deposition by pulsed laser deposition (PLD). Introduction of hydrogen gas along with oxygen gas directly into the vacuum chamber is effective but poses safety risks. Adjusting the gate valve to introduce water vapor into the chamber is a safer alternative but is complicated and impractical. In contrast, using an In(OH)$_3$ ceramic target enables direct hydrogen incorporation into the film, providing a safe, simple, and effective approach.

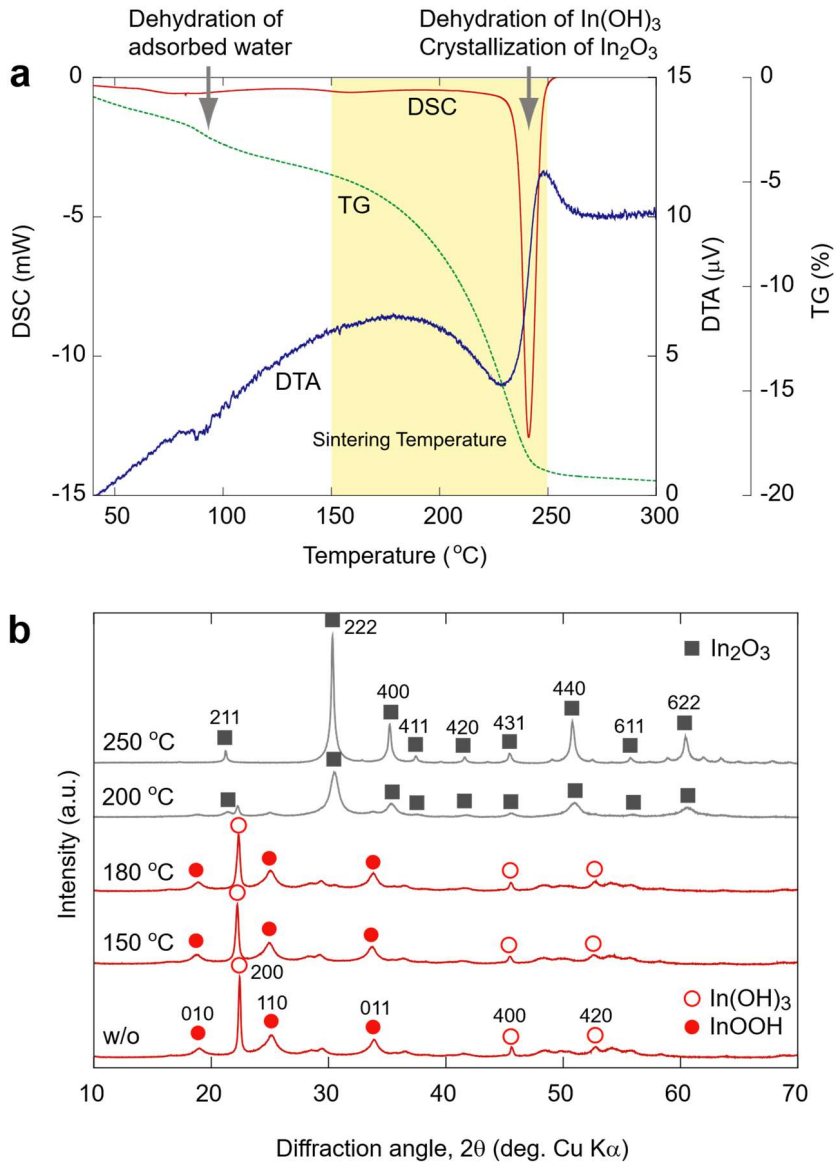

**Figure 2. Determination of sintering temperature of In(OH)₃.** (a) DSC and TG-DTA curves of In(OH)₃ powder. A small endothermic DTA peak around 90 °C corresponds to the vaporization of surface-adsorbed water molecules. A broader endothermic DTA peak around 230 °C, accompanied by gradual weight loss, indicates dehydration of In(OH)₃. An abrupt exothermic DTA peak around 250 °C corresponds to the crystallization of In₂O₃. Based on this data, the sintering temperature range for In(OH)₃ ceramic targets was set between 150 − 250 °C. (b) Powder XRD patterns of In(OH)₃ ceramic targets sintered at various temperatres. Below 180 °C, the In(OH)₃ and InOOH phases remain stable. Above 200 °C, diffraction peaks of In₂O₃ emerge and become sharper with increasing temperature, indicating crystal growth of In₂O₃.

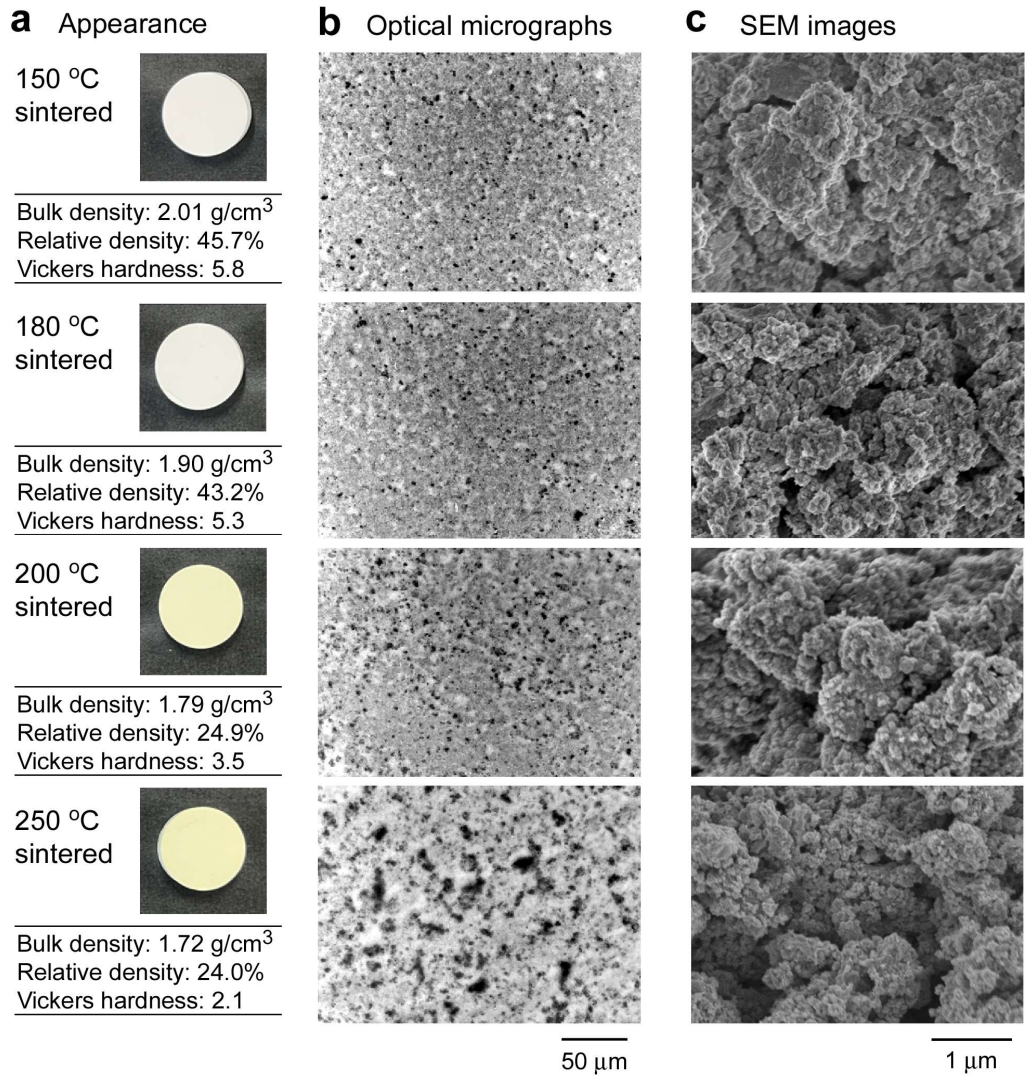

**Figure 3. Microstructure of the resultant In(OH)$_3$ ceramic targets.** (a) Phtograph, bulk density, relative density, and Vickers hardness of the ceramic targets. The relative density of the 150 and 180 °C-sintered targets was calculated relative to theoretical density of In(OH)$_3$, while that of 200 and 250 °C-sintered targets was calculated relative to theoretical density of In$_2$O$_3$. (b) Optical micrographs of the ceramic targets showing an increase in open pores with increasing sintering temperature. (c) SEM images of the ceramic targtes. The grain size appears unchanged across different sintering temperatures.

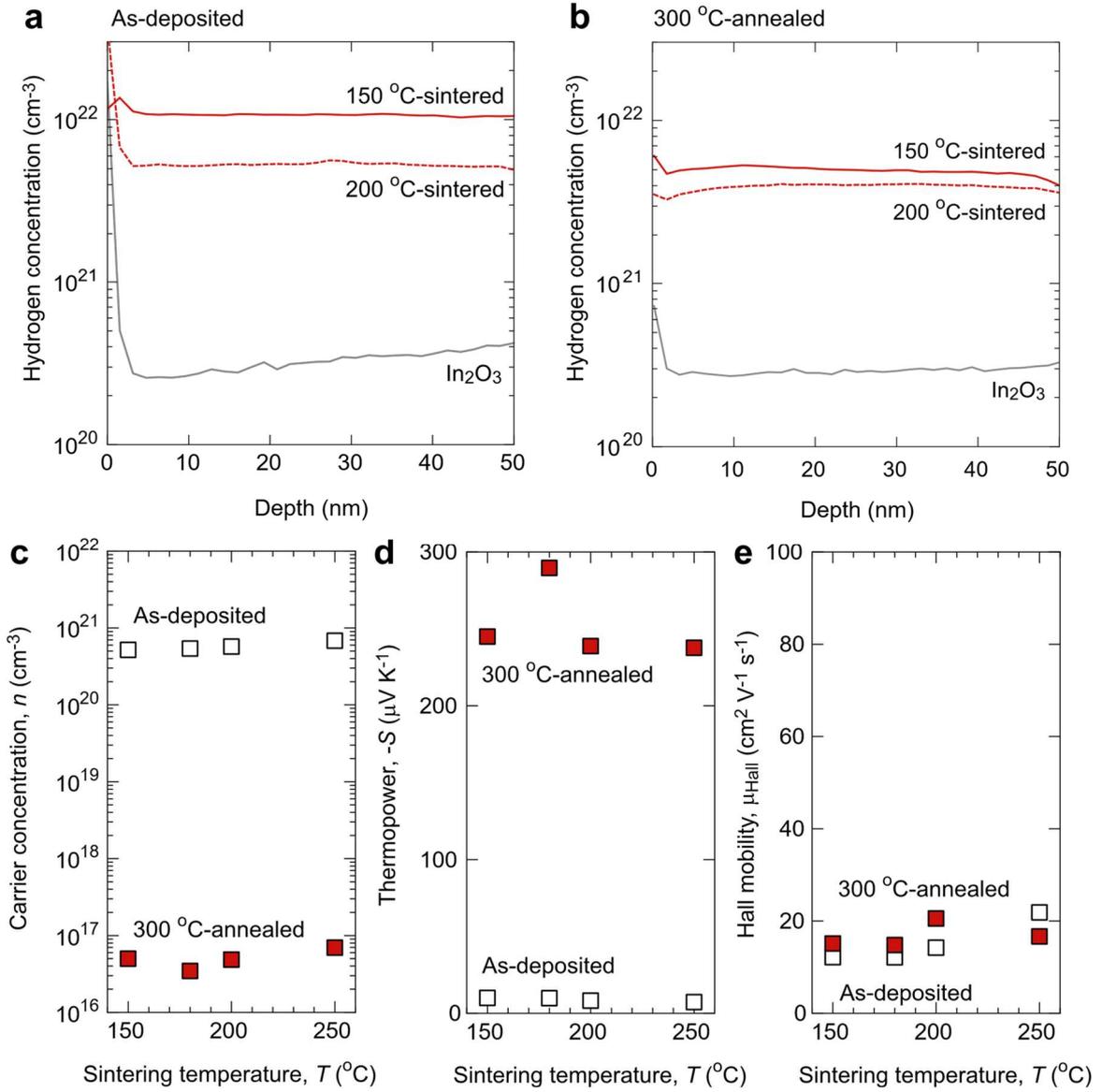

**Figure 4. Evidence of hydrogen incorporation in the as-deposited films.** (a, b) SIMS depth profiles of hydrogen concentration in (a) the as-deposited films and (b) the 300 °C-annealed films. The hydrogen concentration is homogenious across the films (150 °C-sintered In(OH)$_3$ target: ~1 × 10$^{22}$ cm$^{−3}$, 200 °C-sintered In(OH)$_3$ target: ~5 × 10$^{21}$ cm$^{−3}$, dense In$_2$O$_3$ target: ~3 × 10$^{20}$ cm$^{−3}$). (c, d, e) Electron transport properties of the resultant films measured at room temperature.

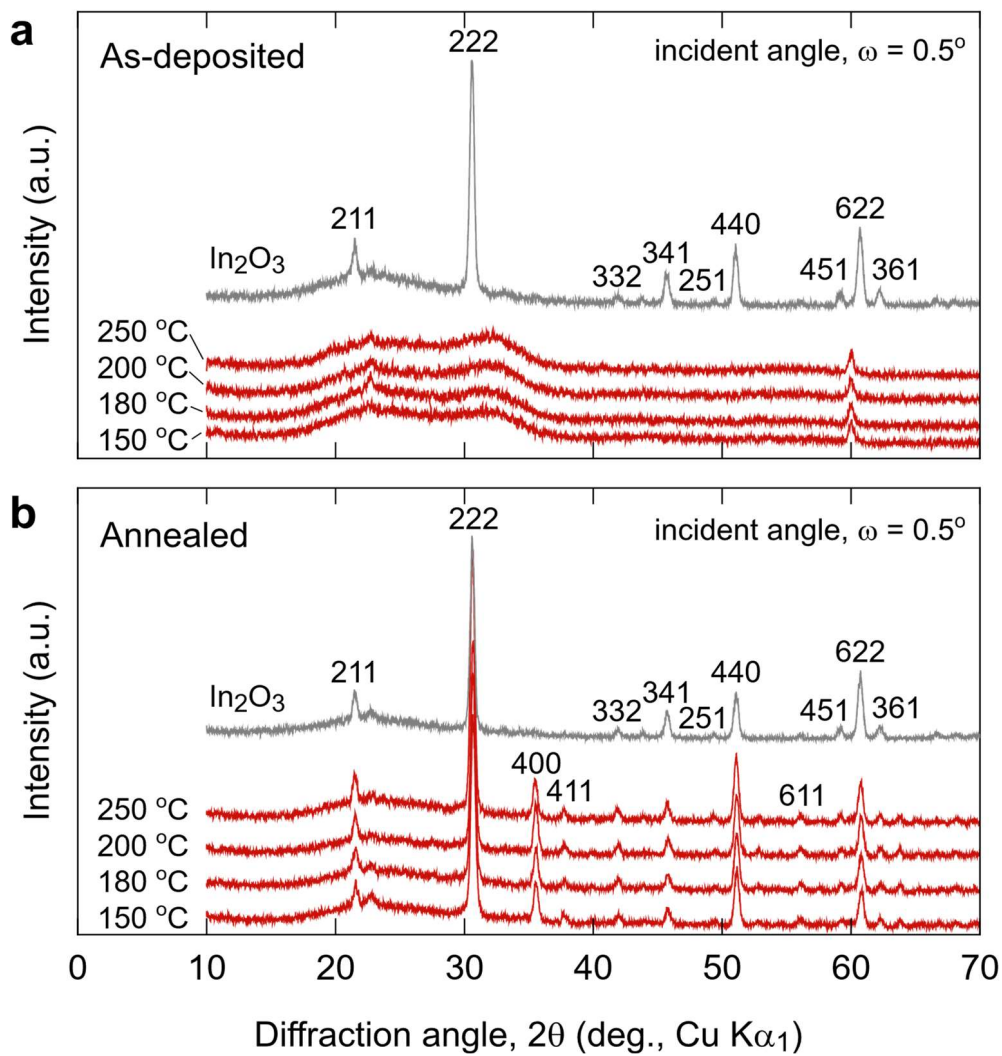

**Figure 5. Grazing incidence XRD patterns of the resultant films deposited using the In(OH)$_3$ ceramic targets.** (a) As-deposited films. A halo pattern around $2\theta = 32°$ indicates an amorphous structure in films deposited using In(OH)$_3$ ceramic targets. In contrast, intense diffraction peaks that indicate randomly oriented polycrystalline are observed in films deposited using a dense In$_2$O$_3$ target. The peak-like feature at $2\theta = 60°$ orginates from the instrument. (b) Films annealed at 300 °C in air. All films exhibit intense diffraction peaks, confirming the formation of randomly oriented polycrystalline structures.

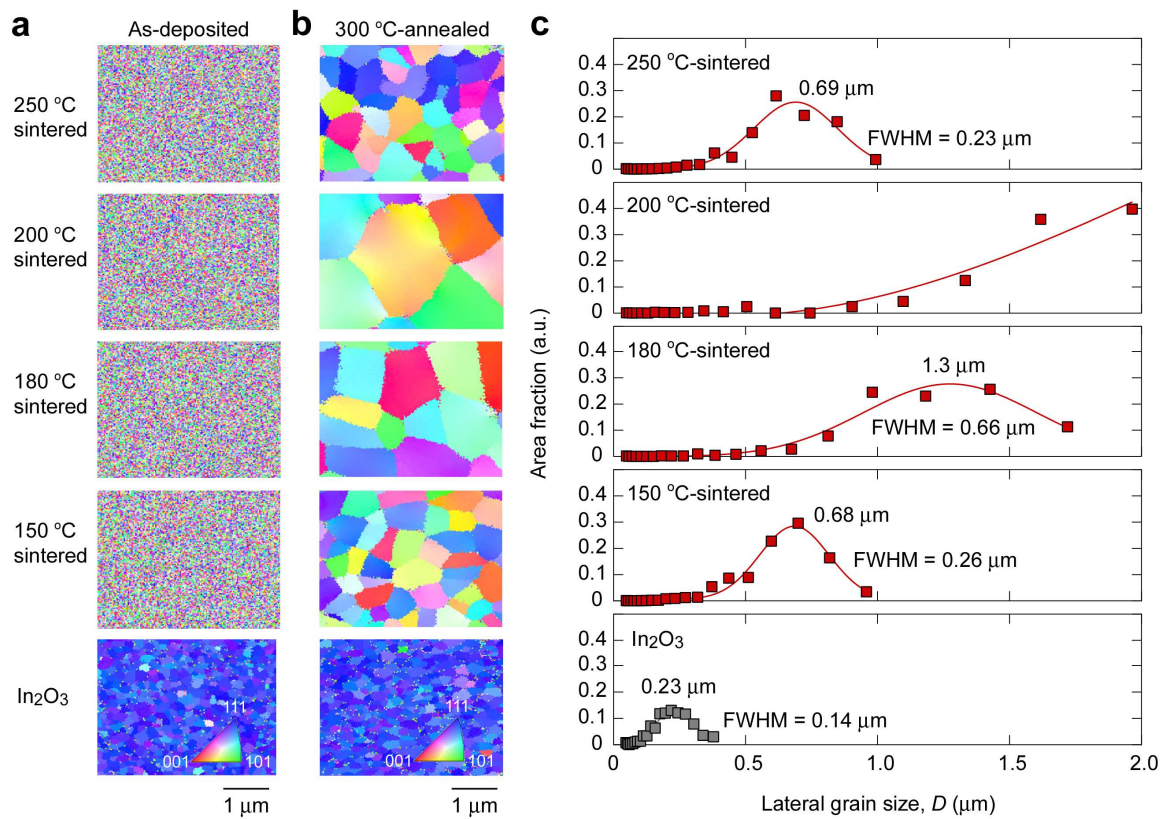

**Figure 6. Lateral grain size of the resultant films deposited using the In(OH)$_3$ ceramic targets.** (a) EBSD images of as-deposited films. Films deposited using In(OH)$_3$ ceramic targets remain amorphous, while those deposited using a dense In$_2$O$_3$ ceramic target exhibit a small-grain polycrystalline structure. (b) EBSD images of 300 °C-annealed films. Large grains appear in films deposited using In(OH)$_3$ ceramic targets, with the largest grain size observed in the 200 °C-sintered target. In contrast, films deposited using the dense In$_2$O$_3$ ceramic target show minimal grain growth after annealing. (c) Area fraction of the lateral grain size, extracted from the EBSD images. The solid lines represent Gaussion peak fitting results.

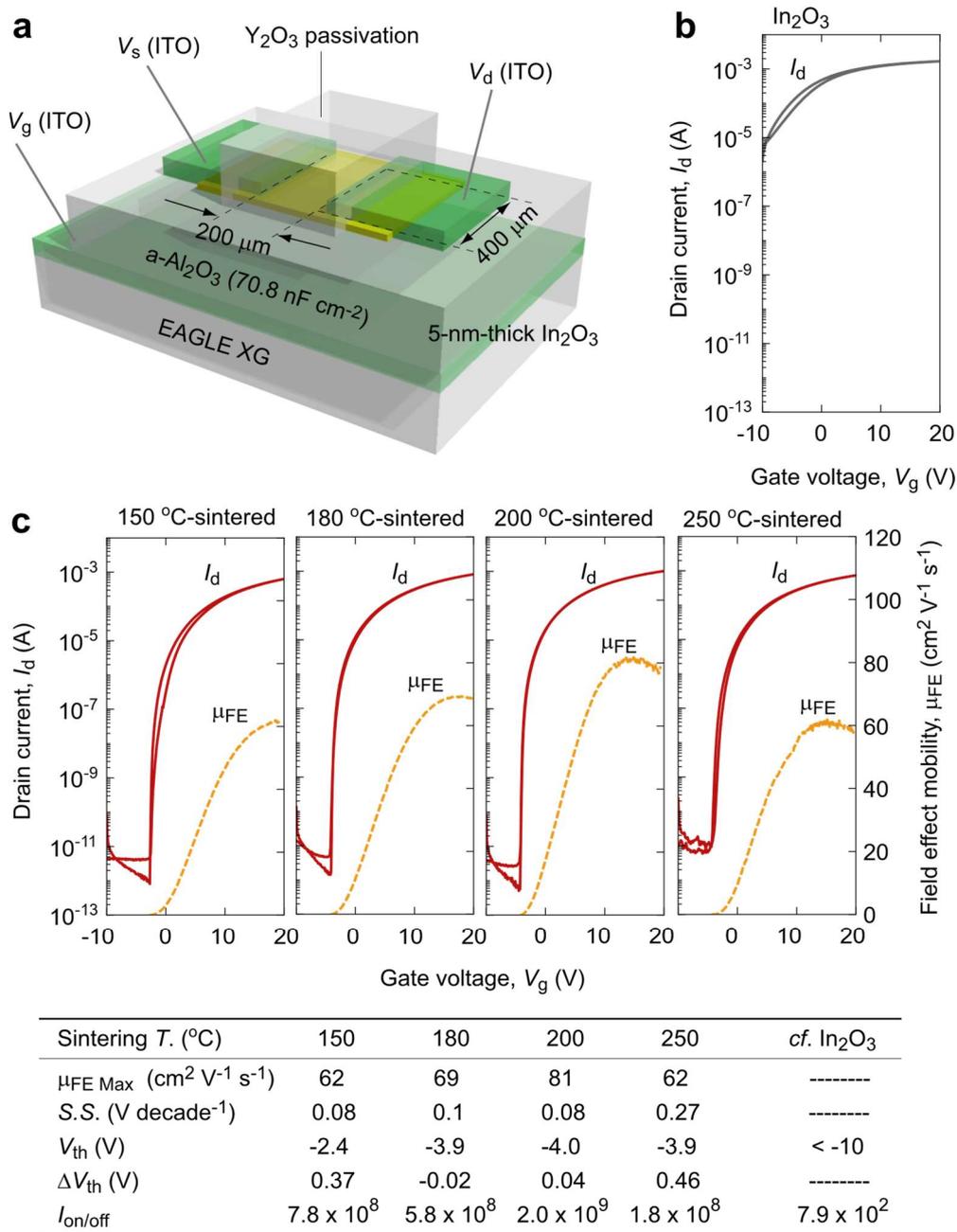

**Figure 7. Transistor characteristics of In$_2$O$_3$-based TFTs using the In(OH)$_3$ ceramic targets.** (a) Schematic device structure of the TFTs. (b) The data for the TFT using a dense In$_2$O$_3$ ceramic target is also shown for comparison. (c) Transfer characteristics of the resultant TFTs. All TFTs show good switching characteristics with high $\mu_{FE}$ >60 cm$^2$ V$^{-1}$ s$^{-1}$ except the dense In$_2$O$_3$ target one. Note that the TFT with 200 °C-sintered In(OH)$_3$ ceramic target shows rather high $\mu_{FE}$ of ~ 81 cm$^2$ V$^{-1}$ s$^{-1}$, probably due to the largest grain size among them.

# Indium Hydroxide Ceramic Targets: A Breakthrough in High-Mobility Thin-Film Transistor Technology

Hikaru Sadahira[1,#] Prashant R. Ghediya[2,#] Hyeonjun Kong[1,#] Akira Miura[3], Yasutaka Matsuo[2], Hiromichi Ohta[2,*] and Yusaku Magari[2,*]

[1] *Graduate School of Information Science and Technology, Hokkaido University, N14W9, Sapporo 060-0814, Japan*
[2] *Research Institute for Electronic Science, Hokkaido University, N20W10, Sapporo 001-0020, Japan*
[3] *Graduate School of Engineering, Hokkaido University, N13W8, Kita, Sapporo 060-8628, Japan*

\* hiromichi.ohta@es.hokudai.ac.jp, magari.yusaku@es.hokudai.ac.jp
#Equally contributed to this work

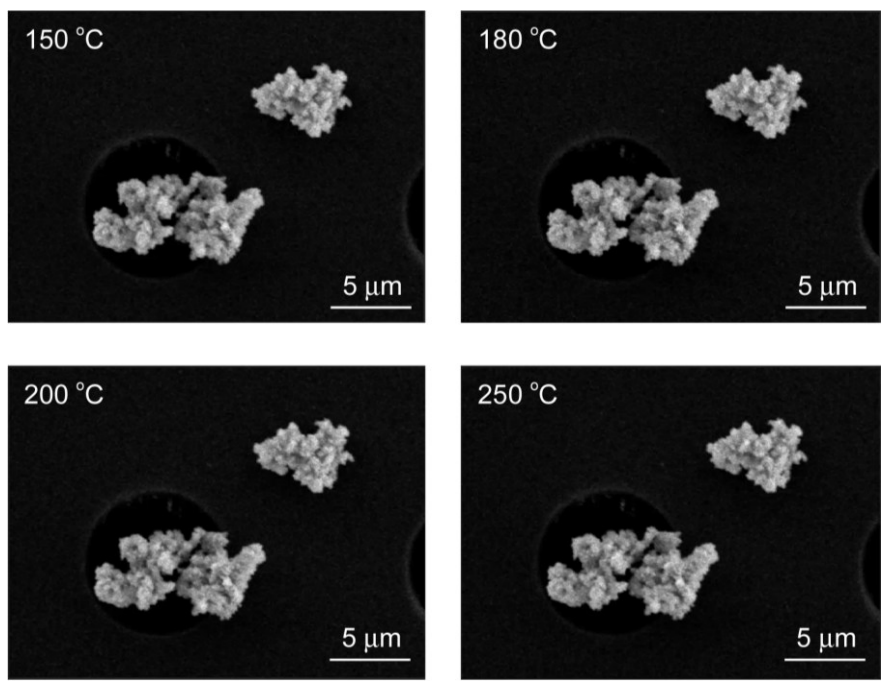

**Figure S1.** High-temperature SEM images of In(OH)$_3$ powder. The overall shape of the In(OH)$_3$ powder did not change up to 250 °C.

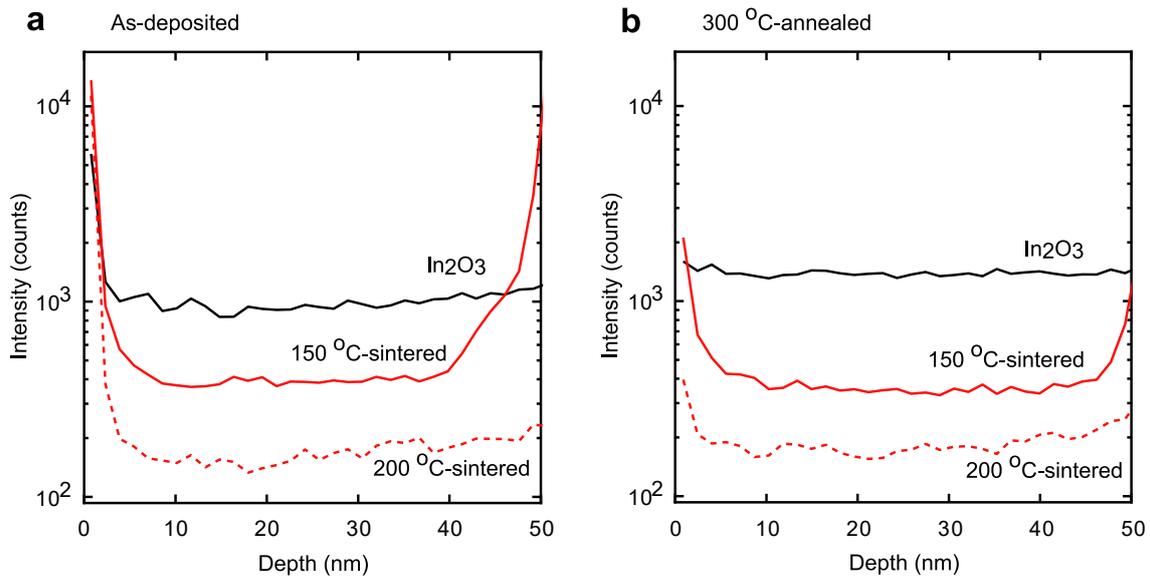

**Figure S2.** SIMS depth profiles of carbon concentration in (a) the as-deposited films and (b) the 300 °C-annealed films. The $In_2O_3$ films deposited using the $In(OH)_3$ ceramic target contained fewer carbon impurities than those deposited using a conventional dense $In_2O_3$ ceramic target, both before and after annealing. In the $In(OH)_3$ ceramic target sintered at 150 °C, the increased signal intensity in the bulk region is due to the SIMS matrix effect occurring at the underlying $SiO_2$ interface.

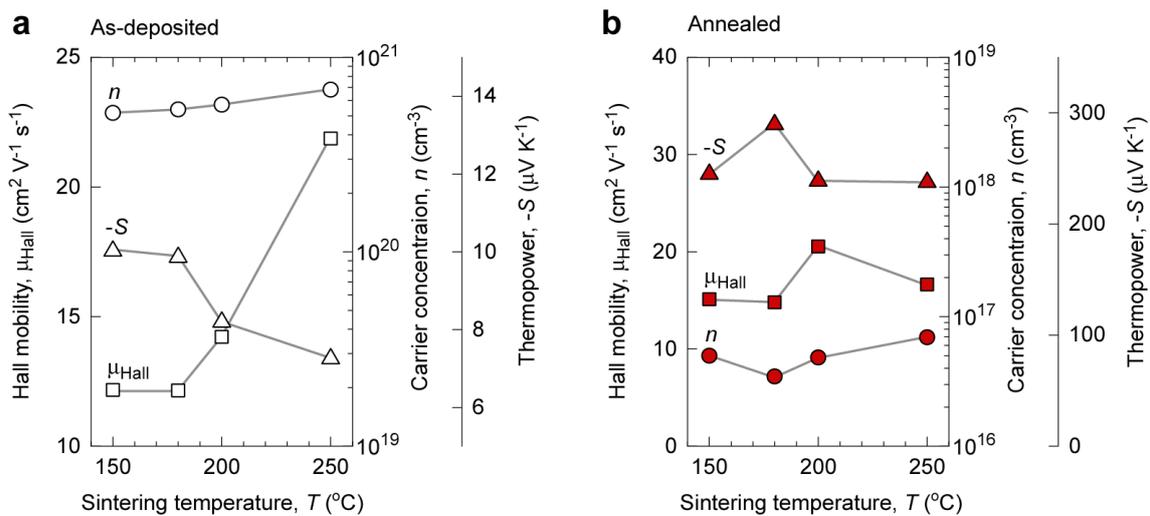

**Figure S3.** Electron transport properties of the resultant films measured at room temperature. The figure compares the carrier concentration ($n$), Hall mobility ($\mu_{Hall}$), and thermopower ($S$) between (a) as-deposited and (b) 300 °C-annealed films.

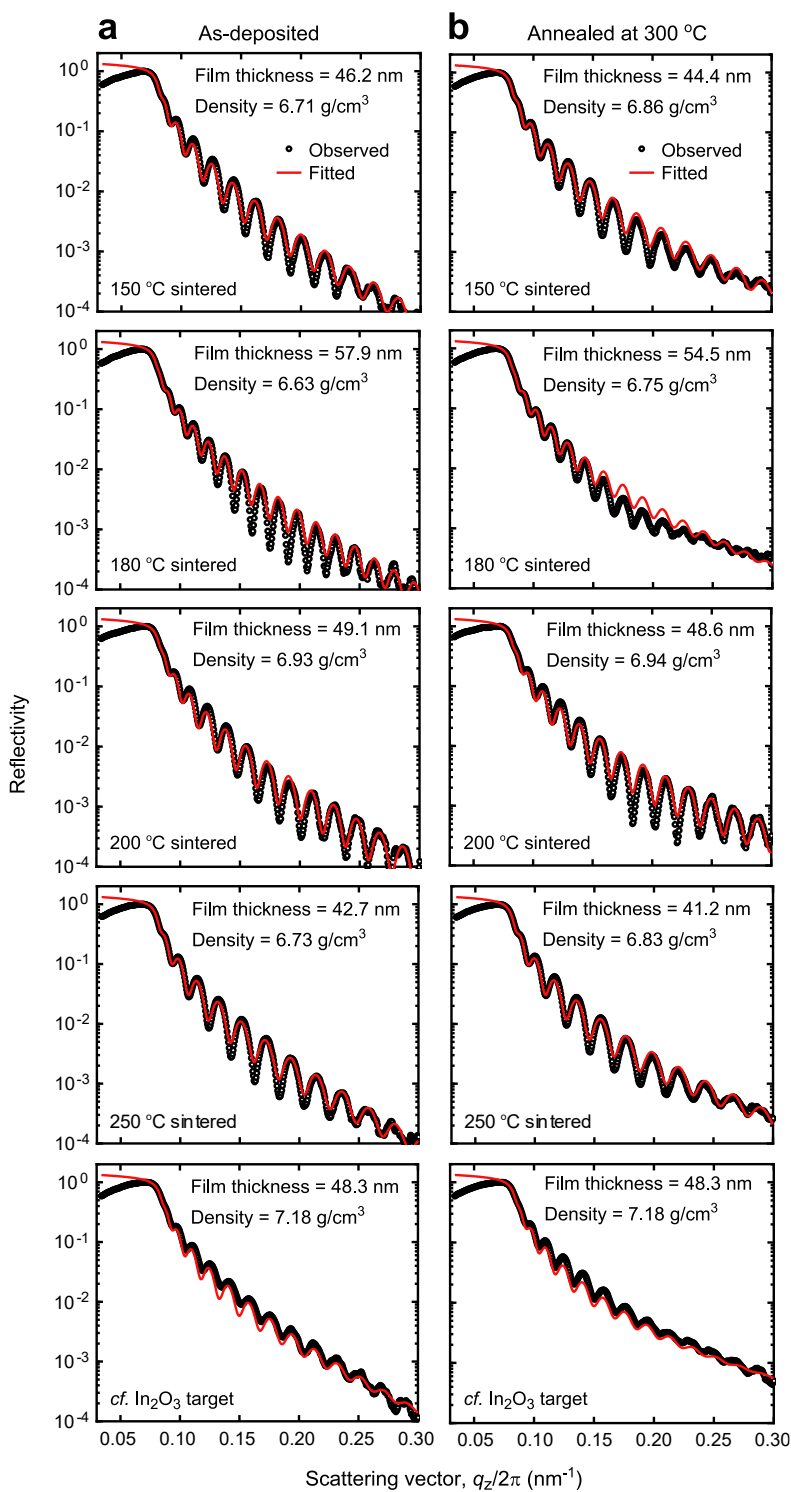

**Figure S4.** Bulk density of the films using In(OH)$_3$ ceramic targets. X-ray reflectivity (XRR) measurements of films. (a) As-deposited and (b) 300 °C-annealed. After annealing, the film density increased for those deposited from In(OH)$_3$ ceramic targets.

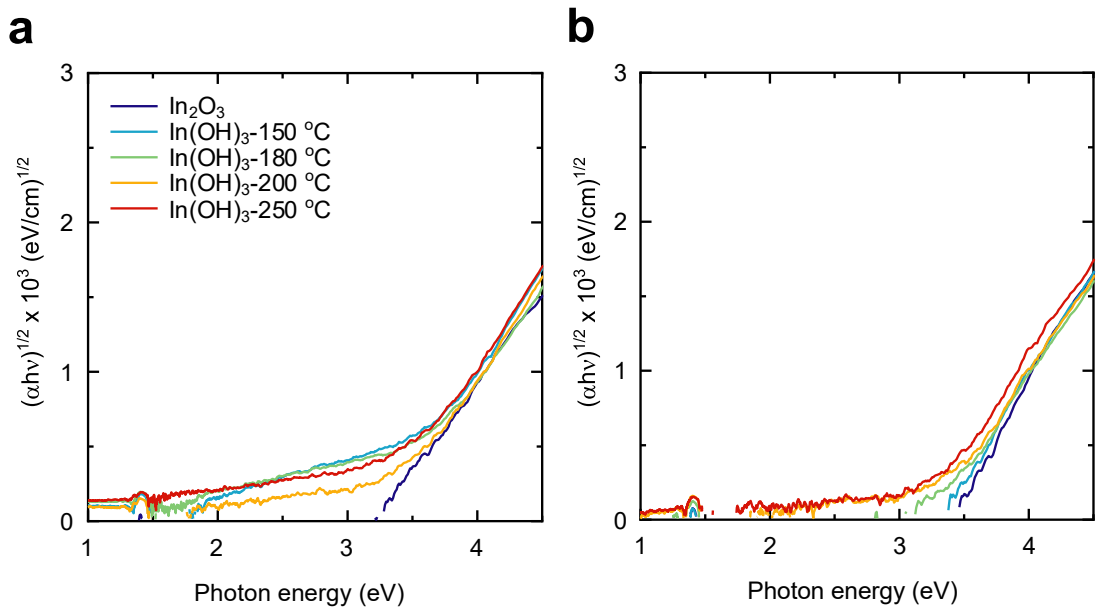

**Figure S5.** Optical bandgap of the films using In(OH)$_3$ ceramic targets. Optical bandgap values calculated from optical absorption measurements for the films. The bandgap was determined using Tauc plots derived from the absorption spectra. (a) The as-deposited films show bandgaps of 3.25 eV, 3.12 eV, 3.16 eV, 3.23 eV and 3.13 eV for the In$_2$O$_3$ ceramic target, 150 °C, 180 °C, 200 °C, and 250 °C-sintered In(OH)$_3$ target, respectively. (b) The 300 °C-annealed films show bandgaps of 3.37 eV, 3.34 eV, 3.27 eV, 3.29 eV and 3.23 eV for the In$_2$O$_3$ ceramic target, 150 °C, 180 °C, 200 °C, and 250 °C-sintered In(OH)$_3$ target, respectively.

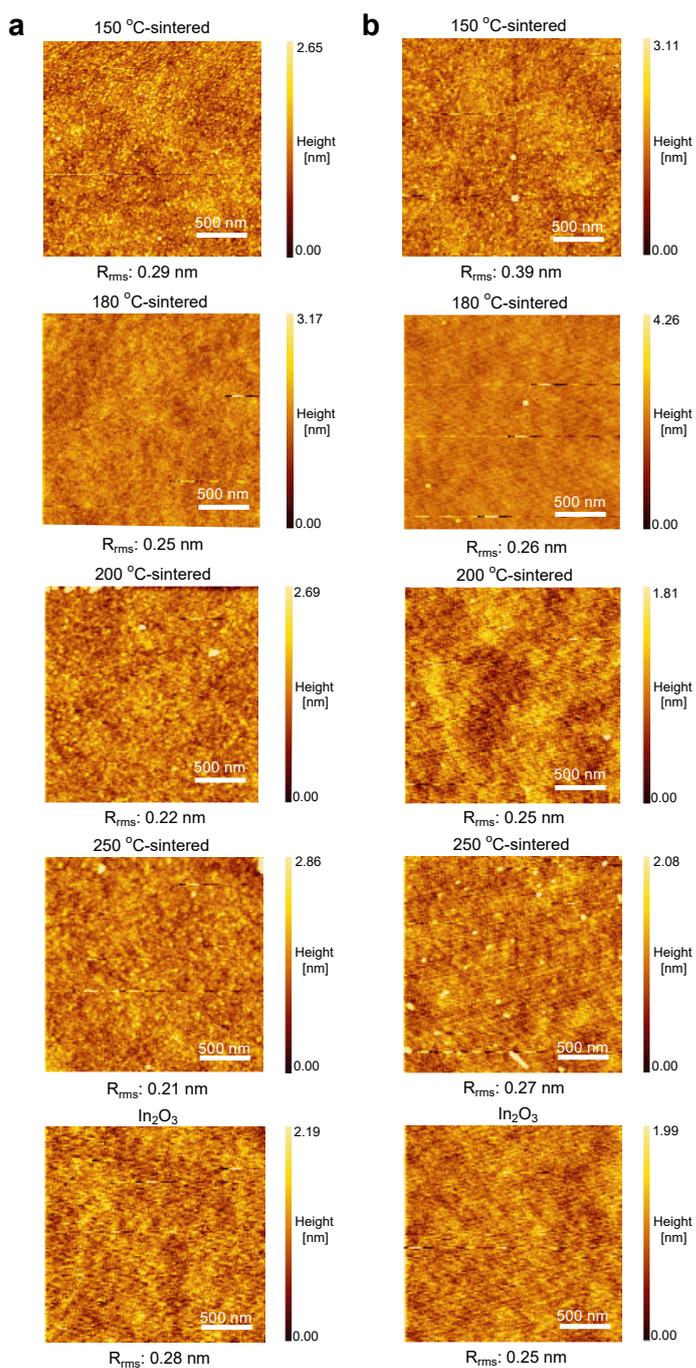

**Figure S6.** Topographic AFM images of the films using In(OH)$_3$ ceramic targets. Atomic force microscopy (AFM) images of (a) as-deposited and (b) 300 °C-annealed films deposited from In(OH)$_3$ and In$_2$O$_3$ ceramic targets. The surface morphology remains relatively smooth before and after annealing, with low root-mean-square roughness ($R_{rms}$). The scale bar represents 500 nm.

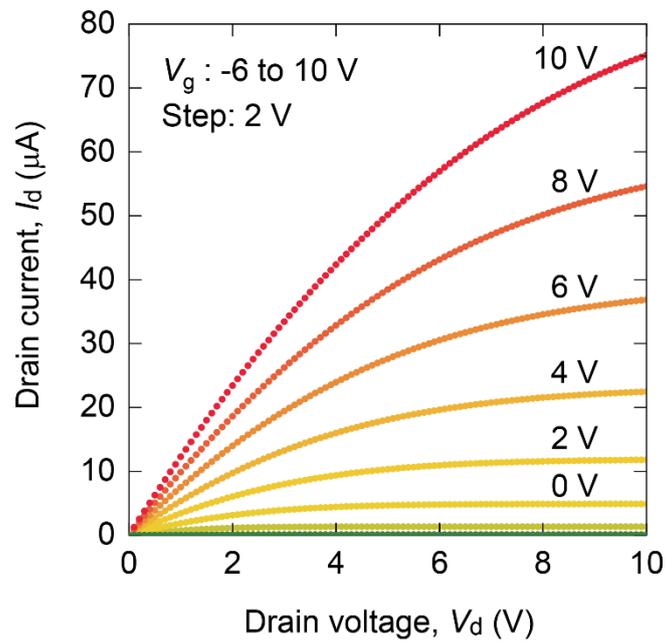

**Figure S7.** Output characteristics of $In_2O_3$-based TFTs using the $In(OH)_3$ ceramic targets. The $I_d$ showed pinch-off behavior and current saturation characteristics at higher $V_d$.

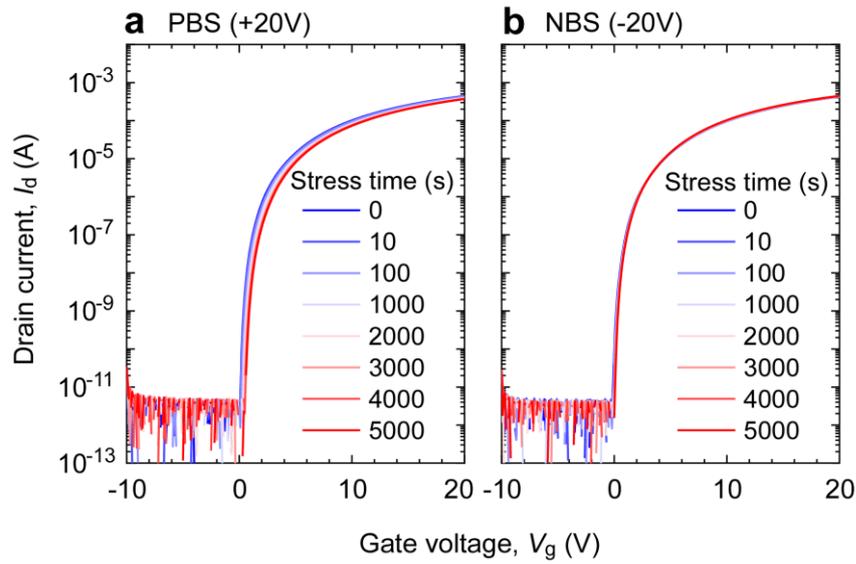

**Figure S8.** Bias stress stability test of the resultant $In_2O_3$ TFTs fabricated using the $In(OH)_3$ ceramic target. (a) Positive bias stress (PBS: $V_g = +20$ V), (b) negative bias stress (NBS: $V_g = -20$ V). The threshold voltage ($V_{th}$) shifts after 5000 s of PBS and 5000 s of NBS were negligible, indicating the high reliability of the $In_2O_3$ TFT fablicated using $In(OH)_3$ ceramic target.